\documentclass{article}

\usepackage[final]{neurips_2022}

\usepackage[utf8]{inputenc} % allow utf-8 input
\usepackage[T1]{fontenc}    % use 8-bit T1 fonts
\usepackage{hyperref}       % hyperlinks
\usepackage{url}            % simple URL typesetting
\usepackage{booktabs}       % professional-quality tables
\usepackage{amsfonts}       % blackboard math symbols
\usepackage{nicefrac}       % compact symbols for 1/2, etc.
\usepackage{microtype}      % microtypography
\usepackage{xcolor}         % colors

\title{Redefining Relationships in Music}

\author{%
  Christian Detweiler\thanks{c.a.detweiler@hhs.nl}\\
  The Hague University of Applied Sciences\\
  \And
  Beth Coleman \\
  University of Toronto \\
  \AND
   Fernando Diaz \\
   Google \\
  \And
  Lieke Dom \\
  Google \\
\And
  Chris Donahue \\
  Google \\
\And
  Jesse Engel \\
  Google \\
\And
  Cheng-Zhi Anna Huang \\
  Google \\
\And
  Larry James \\
  Google \\
\And
  Ethan Manilow \\
  Google \\
\And
  Amanda McCroskery \\
  Google \\
\And
   Kyle Pedersen \\
  Google \\
\And
  Pamela Peter-Agbia \\
  Google \\
\And
  Negar Rostamzadeh \\
  Google \\
\And
  Robert Thomas \\
\And
  Marco Zamarato \\
  Google \\
  \And
  Bendert Zevenbergen \\
  Google \\
}

\begin{document}

\maketitle

\section{Introduction}
AI tools increasingly shape how we discover, make and experience music. While these tools can have the potential to empower creativity, they may fundamentally redefine relationships between stakeholders, to the benefit of some and the detriment of others. In this position paper, we argue that these tools will fundamentally reshape our music culture, with profound effects (for better and for worse) on creators, consumers and the commercial enterprises that often connect them. By paying careful attention to emerging Music AI technologies and developments in other creative domains and understanding the implications, people working in this space could decrease the possible negative impacts on the practice, consumption and meaning of music. Given that many of these technologies are already available, there is some urgency in conducting analyses of these technologies now. It is important that people developing and working with these tools address these issues now to help guide their evolution to be equitable and empower creativity. We identify some potential risks and opportunities associated with existing and forthcoming AI tools for music, though more work is needed to identify concrete actions which leverage the opportunities while mitigating risks. 

\subsection{Positionality statement}
This position paper is based on discussions that emerged during a two-day workshop on sociotechnical considerations of AI in music technology. Given the scope of the problem considered, we would like to acknowledge our technical affiliations and academic backgrounds are relatively narrow, with most members working in a large tech corporation and sharing affiliation with several select academic institutions in North America and Western Europe.

\section{Technologies considered}
In the workshop we selected a variety of music tools that make use of some form of AI as a starting point for discussions about broader implications of this technology. The idea was not to "call out" any shortcomings of these particular tools, but to ground our discussions in experiences with concrete tools. The discussions surfaced a number of common attributes. The selected tools differ in the amount of control or agency they offer users. Whereas some of the tools generate a complete audio file within a few clicks, others allow users to modify and run code to influence how the tool generates music. Tools differ with regard to the visibility of model provenance. For example, in some tools, users can upload MIDI files to influence music generation or create models using their own audio. Other tools do not explicitly reference models or source material in the functions they offer users. 

\section{Actors affected}
The tools we examined target a variety of users. They provide non-musicians with novice-friendly ways of making music. Professional creatives (e.g., video producers) can also use some of these tools to cheaply and easily create (royalty-free) music to fit their content. New "AI artists" can use tools to create new forms of art and in doing so often push tools' limits and help improve them as can be seen, for example, in the work of Portrait XO. In some cases, large corporations create these tools and use them to provide music as a service. Although consumers might not encounter these tools directly, they could hear their output as the (generated) soundtrack to a video or podcast, or new types of music created by new AI artists. In some cases, consumers might become creators themselves, as can be seen on the Jukebox Discord server (Jukebox, n.d.). These tools rely on training data. This could come from living musicians (as "style" data), but also from non-living recorded musicians. This raises questions about consent for being included in a dataset, which projects such as Spawning are beginning to address by building tools to allow artists to opt in or opt out of their data being included in the training of large AI models and set permissions on how their style and likeness is used (Dryhurst et al., 2022). 

\section{Opportunities and risks}
AI tools are changing how musicians, cultural producers, and listeners relate to each other and to musical expression more broadly. In our discussions, we identified a number of opportunities and risks associated with existing and emerging tools. AI tools create possibilities for new types of roles and relationships in music. Configurations of performer-instrument-audience or composer-score-performer are shifting or breaking down with the use of AI in music (Magnusson, 2019). Such changes in musical practices can alter established relationship dynamics by expanding the notion of authorship, potentially diluting it, and correspondingly the connection between the creator and the listener, and trust in the shared intention of using this music for a shared societal purpose. There are already examples of artists pushing the limits of these tools to create new forms of musical expression and art more broadly, such as Holly Herndon's Holly+ (Herndon, 2019). Music making could take on new meanings if these tools incentivize creation of input for new models. 

\subsection*{Credit, compensation or exposure in or from these tools}
That said, AI tools in music redefine the relationships between certain stakeholders and value captured by certain (other) stakeholders. There are potential risks to how these technologies will both borrow from and impact existing musicians, and transfer captured value to other stakeholders (Drott, 2021). Most of the tools we analyzed rely on large amounts of recorded audio to create the models used to generate new audio. Once these systems are trained, determining how much the system relies on a given piece of source material is currently an underexplored topic. Therefore, what "influences" the system uses when it generates a new piece is difficult to know. As a result, few tools are transparent about what source material they rely on for specific outputs. The musicians who "bring in the sounds" rarely receive credit, compensation or exposure in or from these tools, though exceptions exist. For example, CoSo by Splice combines bits of existing music, making attribution clear and allowing musicians to receive money and credit for their work (Splice, 2022). 

\subsection*{Homogenization of the landscape of music}
A further risk is that these tools homogenize the landscape of music. The tools we examined encapsulate specific ideas about what music is (e.g., music as a product to be consumed rather than as “musicking” (Small, 1998), specific economic models, musical aesthetics, values and music theories of specific cultural contexts. These affect the way people conceive, produce and master musical work (Magnusson, 2019). “Automated aesthetic judgments” can be used to shape the way media sound (Sterne and Razlogova, 2021). The volume of music made with AI tools will potentially outpace that of music made without such tools. If AI tools lack diverse outputs, the landscape of music might homogenize quickly, both through more overt influence, such as autotune, and more subtle constraints on what is generated. 
\subsection*{Changing notions of musicianship, mastery and skill}
These tools can also change notions of musicianship, mastery and skill (Tahiroğlu, 2021). Creating music traditionally requires honing a skill over time (Gurevich, 2014). Some AI tools have a lower barrier to entry, which can increase access to novices, but also potentially diminish the perceived value of musical practice, reducing the public conception of music to a commodity and musicianship to an automatable task (Morreale, 2021). This could impact musicians’ ability to economically sustain themselves while they develop and practice their craft. Alternatively, as more consumers engage with music creation, it could drive newfound levels of appreciation for the process of creation and those who dedicate their life to it, similar to how widespread access to digital audio workstations has increased the musical literacy of the public and popular video games such as Guitar Hero have created new waves of fans for technically virtuosic musicians featured in the games.
\section{Open questions for the community}
Based on these issues, we identify six themes that merit further attention.
\begin{itemize}

\item Centering artists in the process of developing these tools, and in the tools themselves
\item Creating systems that empower more creativity than they borrow
\item Improving interpretability of these systems and their (intermediate) outputs, and giving credit, exposure and compensation to artists where it is due
\item Questioning what parts of the process are being automated and to whose benefit
\item Centering new music technologies in ways that are responsive to non-Western cultures
\item Studying the broader and long term impact of these tools
\end{itemize}

\section{Conclusion}
To leverage opportunities and mitigate risks of AI tools in music, we urgently need to study these tools and their potential impact. In this paper, we have consolidated a number of issues that need further work. The discussions captured here really only scratch the surface and more work is needed to come up with frameworks and specific recommendations for how to proceed in light of the risks and opportunities identified here. 
\begin{ack}

This work has been supported by Google through an unrestricted grant. 

\end{ack}

\section*{References}

\small 

Drott, E. (2021). Copyright, compensation, and commons in the music AI industry. Creative Industries Journal, {\bf 14}(2), 190-207

Dryhurst, M., Herndon, H., Hoepner, P. and Meyer, J. (2022). {\it Spawning.} \url{https://spawning.ai/}

Gurevich, M. (2014). Skill in Interactive Digital Music Systems. In K. Collins, B. Kapralos, and H. Tessler (Eds.), {\it The Oxford Handbook of Interactive Audio.} Oxford: Oxford University Press.

Herndon, H. (2021). {\it Holly+.} \url{https://holly.plus/}

Jukebox. (n.d.). {\it Jukebox} [Discord server]. Discord. \url{https://discord.me/jukeboxmusic/} 

Magnusson, T. (2019). {\it Sonic writing: technologies of material, symbolic, and signal inscriptions.} Bloomsbury Academic.

Morreale, F. (2021). Where Does the Buck Stop? Ethical and Political Issues with AI in Music Creation. {\it Transactions of the International Society for Music Information Retrieval}, {\bf 4}(1), 105–113.

Small, C. (1998). {\it Musicking: The meanings of performing and listening.} Wesleyan University Press.

Splice. (2022). {\it CoSo by Splice.} (Version 1.3.7) [Mobile App]. Google Play. \url{https://play.google.com/store/apps/details?id=com.splice.stacksmobile&hl=en&gl=US1} 

Sterne, J. and Razlogova, E. (2021) Tuning sound for infrastructures: artificial intelligence, automation, and the cultural politics of audio mastering, {\it Cultural Studies}, {\bf 35}(4-5), 750-770

Tahiroğlu, K. (2021). Ever-shifting roles in building, composing and performing with digital musical instruments, {\it Journal of New Music Research}, {\bf 50}(2), 155-164

\end{document}